\documentclass[11pt]{article}
\usepackage{amsmath,amssymb}
\usepackage[margin=1in]{geometry}
\usepackage{hyperref}
\usepackage{booktabs}
\usepackage{graphicx}
\usepackage{subcaption}
\usepackage{float}
\usepackage{xcolor}
\usepackage{natbib}

\title{Modeling the Joint Distribution of Box Score Statistics: \\
A Gaussian Copula Approach}

\author{
    Kenny Watts \\
    University of Virginia \\
    \texttt{xen5ek@virginia.edu}
}

\date{November 2026}

\begin{document}

\maketitle

\begin{abstract}
What is the probability that LeBron James records at least 27 points, 7 rebounds, and 7 assists tonight?\textsuperscript{\ref{app:lebron}} While an interesting statistical question by itself, the rise of sports betting and multi-leg parlays has made answering this question increasingly relevant. However, coming up with an estimated likelihood is tricky, since Box Score statistics are correlated. \\
We estimate the joint probability that a player reaches a combination of thresholds in points, assists, and rebounds using a Gaussian copula, fit separately for each player using only their previous games. We compare it with two benchmarks: the product of the empirical marginal probabilities (i.e. assuming independence) and the empirical frequency of the joint event. Across $\sim 22,000$ player games, the Gaussian copula performs on par with both benchmarks at lower threshold levels, with gains at high joint exceedance levels. Because it is generative, the model also allows for probabilistic calculations for combinations that a player has rarely or never produced.
\end{abstract}

\section{Introduction}

Same-game parlays allow bettors to combine several outcomes from a single game into one wager. In basketball these legs often involve player box score statistics; for example, a bettor may wager that a player records at least 20 points, 5 rebounds, and 5 assists. \citep{whelan2026parlay} has found that the expected loss per dollar wagered on parlays is roughly four times higher than non-parlay bets, with Sports Books making $19.2\%$ of the total stakes placed on parlays. This motivates a statistical framework for modeling the joint likelihood of these events.

\begin{figure}[H]
  \centering
  \includegraphics[width=0.5\textwidth]{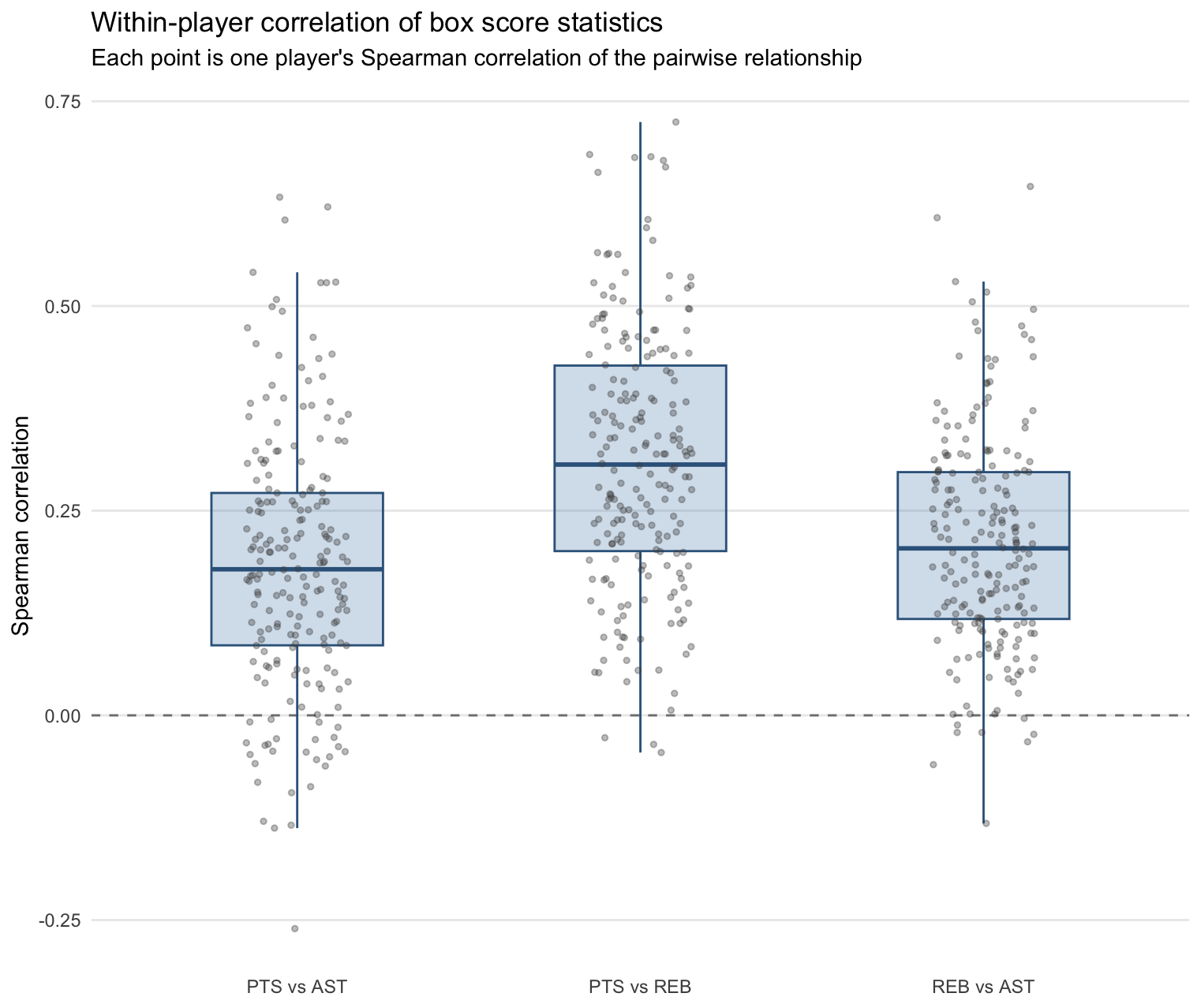}
  \caption{Within player correlation between Box Score Statistics}
  \label{fig:correlation-plot}
\end{figure}

Treating the three box score statistics as independent is a poor approximation: a player who scored a large number of points likely spent more time on the court, accumulating opportunities to record rebounds and assists. Modeling these joint probabilities therefore requires a framework that accounts for inter-variable dependencies, rather than treating each metric in isolation. Figure \ref{fig:correlation-plot} shows that our three box score statistics exhibit mild but positive pairwise correlation.

Copulas provide a natural framework for this problem, since they separate each variable's marginal distribution from the dependence structure that links them \citep{sklar1959fonctions}. Once the marginals and the copula are estimated, we can simulate from the fitted joint distribution and estimate the probability of any combination of variable thresholds, including combinations with little or no historical support. Gaussian copulas were popularized in finance for modeling default correlation \citep{li2000default}, and copula methods have been applied to basketball data before; \citet{durso2023bayesian} used a copula-based Bayesian network to study the determinants of player performance, and \citet{liu2026bayesian} used a copula-based Dirichlet process mixture to find subgroups of similar players. In Soccer, \citep{vanderwurp2020generalised} proposed a copula-based joint regression model for the number of goals scored by each team, allowing scoring rates to depend on the current match state and scoreline. Our objective is to model the joint distribution of points, rebounds, and assists at the individual player level.\\

In this paper, we model this joint distribution using a Gaussian copula, fitting the model in a strictly out-of-sample setting by re-estimating it before every game. We evaluate joint exceedance forecasts across 27 combinations of player-relative thresholds with Log Loss as our primary loss function, comparing the Gaussian copula against independence and empirical-frequency benchmarks. The results show that the benefits of modeling dependence are concentrated in combinations where all three thresholds are high. The model uses only a player's own game history and does not account for opponents, minutes, or pace. It therefore provides a baseline for assessing the predictive value of dependence modeling rather than a complete model pipeline.

\section{Gaussian Copulas}

A copula is a multivariate distribution function on the unit hypercube $[0,1]^d$ whose univariate margins are uniform. Sklar's theorem \citep{sklar1959fonctions} states that for any $d$-dimensional joint distribution function $F$ with marginals $F_1, \dots, F_d$, there exists a copula $C$ such that
\begin{equation}
F(x_1, \dots, x_d) = C\big(F_1(x_1), \dots, F_d(x_d)\big),
\end{equation}
and $C$ is unique when the margins are continuous. The marginals describe each variable by itself, and the copula carries the information about their joint behavior, meaning the two can be specified and estimated separately. When the margins are discrete, as with box score counts, $C$ is not unique; we return to this in Section~\ref{sec:design}.

Let $\Phi$ denote the standard normal distribution function and $\Phi^{-1}$ its quantile function. The Gaussian copula is the dependence structure of a multivariate normal distribution,
\begin{equation}
C_{\mathbf{R}}(u_1, \dots, u_d) = \Phi_{\mathbf{R}}\big(\Phi^{-1}(u_1), \dots, \Phi^{-1}(u_d)\big),
\end{equation}
where $\Phi_{\mathbf{R}}$ is the $d$-variate standard normal distribution function with correlation matrix $\mathbf{R}$. The correlation matrix fully determines the dependence. The Gaussian copula captures dependence through a single correlation matrix, allowing each
pair of statistics its own strength of association while retaining flexibility in their marginals.

\section{Estimation}

For each player we model $\mathbf{X} = (X_{\text{pts}}, X_{\text{reb}}, X_{\text{ast}})$ with a Gaussian copula and empirical marginals. To mimic a genuine forecasting setting, the model is re-estimated before every game using only past games: for game $t \geq 21$, the training set consists of the $n = t - 1$ preceding games $\mathbf{x}_1, \dots, \mathbf{x}_n$.

\paragraph{Marginals.}
For $j \in \{\text{pts}, \text{reb}, \text{ast}\}$, the marginal $\hat{F}_j$ is the empirical distribution function of the player's training games, with quantile function $\hat{Q}_j = \hat{F}_j^{-1}$ given by the linearly interpolated sample quantile.

\paragraph{Dependence.}
Each observation is mapped to the unit interval through its rank-based pseudo-observation and then to the normal scale,
\begin{equation}
\hat{u}_{sj} = \frac{r_{sj}}{n + 1}, \qquad z_{sj} = \Phi^{-1}(\hat{u}_{sj}), \qquad s = 1, \dots, n,
\end{equation}
where $r_{sj}$ is the rank of $x_{sj}$ among the training games, with average ranks for ties. The copula's correlation matrix \(\hat{R}\) is estimated as the sample correlation matrix of the normal scores \(z_s = (z_{s,\mathrm{pts}}, z_{s,\mathrm{reb}}, z_{s,\mathrm{ast}})\). We use the `nearPD` procedure from the R package `Matrix` \citep{matrixpackage} to obtain the nearest positive definite correlation matrix, enforcing a unit diagonal. The fitted joint distribution is
\begin{equation}
\hat{F}(x_{\text{pts}}, x_{\text{reb}}, x_{\text{ast}}) = \Phi_{\hat{\mathbf{R}}}\big(\Phi^{-1}(\hat{F}_{\text{pts}}(x_{\text{pts}})),\, \Phi^{-1}(\hat{F}_{\text{reb}}(x_{\text{reb}})),\, \Phi^{-1}(\hat{F}_{\text{ast}}(x_{\text{ast}}))\big).
\end{equation}

\paragraph{Simulation.}
We estimate joint exceedance probabilities by Monte Carlo. For $b = 1, \dots, B$ we draw
\begin{equation}
\mathbf{z}^{(b)} \sim N(\mathbf{0}, \hat{\mathbf{R}}), \qquad u_j^{(b)} = \Phi\big(z_j^{(b)}\big), \qquad x_j^{(b)} = \hat{Q}_j\big(u_j^{(b)}\big),
\end{equation}
giving simulated games $\mathbf{x}^{(b)}$ with the player's own marginals and the estimated dependence. We use $B = 10{,}000$. Fitting and simulation used the \texttt{copula} package in R \citep{copulapackage}.

\section{Evaluation Design}
\label{sec:design}

\subsection{Thresholds and events}

For each player-game, we use the player's training games (at least 20) to set player-specific thresholds. For quartile levels $\boldsymbol{\alpha} = (\alpha_{\text{pts}}, \alpha_{\text{reb}}, \alpha_{\text{ast}})$ with each $\alpha_j \in \{0.25, 0.50, 0.75\}$, giving $3^3 = 27$ combinations, the thresholds are $\tau_j = \hat{Q}_j(\alpha_j)$ and the event of interest is
\begin{equation}
A_{\boldsymbol{\alpha}} = \{X_{\text{pts}} \geq \tau_{\text{pts}},\; X_{\text{reb}} \geq \tau_{\text{reb}},\; X_{\text{ast}} \geq \tau_{\text{ast}}\}.
\end{equation}
For example, the combination (PTS Q50, REB Q25, AST Q75) is the event that the player scores at least his historical median in points, at least his 25th percentile in rebounds, and at least his 75th percentile in assists. Because thresholds are computed from each player's own history, a ``high'' threshold is relative to that player and the events are comparable across players with different statistical profiles. Thresholds are updated along with the model before every game. We score each forecast against the realized outcome $y_t = \mathbf{1}\{\mathbf{x}_t \in A_{\boldsymbol{\alpha}}\}$, where $\mathbf{x}_t$ is the stat line of the game being predicted, resulting in 27 binary outcomes for each player every game.

Although the Gaussian copula can theoretically estimate the likelihood of any combination of thresholds, we limit our analysis to quartiles (Q25, Q50, Q75) to limit the complexity of the prediction space and retain explanatory clarity.

Points, rebounds, and assists take integer values, so their marginal distributions are discrete with frequent ties. This means that the copula is not unique, so \(\hat{F}\) is one of many joint distributions consistent with the marginals. To ensure a like-for-like comparison across models, we define each threshold $\tau_j$ as the empirical quantile of the corresponding statistic in the training data, resulting in integer-valued thresholds.

\subsection{Loss Function}

Forecasts are scored with Log Loss:
\begin{equation}
\ell(p, y) = -\big[y \log p + (1 - y)\log(1 - p)\big],
\end{equation}
With sample sizes as small as 20, the empirical joint exceedance probability can sometimes be exactly zero or one, which Log Loss penalizes heavily if the prediction is incorrect. We therefore applied standard Laplace smoothing, estimating the probability of success as
\[
\hat{p}_{\text{Laplace}} = \frac{k+1}{n+2},
\]
where \(k\) denotes the number of observed successes and \(n\) is the total number of observations. We find this greatly improves the Log Loss performance of the empirical benchmark.

\subsection{Forecasts and benchmarks}

The copula forecast of $\Pr(A_{\boldsymbol{\alpha}})$ is the simulated Monte Carlo estimate
\begin{equation}
\hat{p}^{\text{cop}} = \frac{1}{B} \sum_{b=1}^{B} \mathbf{1}\big\{\mathbf{x}^{(b)} \in A_{\boldsymbol{\alpha}}\big\}.
\end{equation}
The \emph{empirical} benchmark is the Laplace-smoothed frequency of the event in the $n$ training games,
\begin{equation}
\hat{p}^{\text{emp}} =
\frac{
1 + \sum_{s=1}^{n} \mathbf{1}\big\{\mathbf{x}_s \in A_{\boldsymbol{\alpha}}\big\}
}{n+2}.
\end{equation}
The \emph{independence} benchmark multiplies the empirical marginal exceedance frequencies,
\begin{equation}
\hat{p}^{\text{ind}} = \prod_{j=1}^{3} \left[ \frac{1}{n}\sum_{s=1}^{n} \mathbf{1}\big\{x_{sj} \geq \tau_j\big\} \right].
\end{equation}
This captures the marginal distributions but assumes independence. \\

\subsection{Data}

Using the hoopR package \citep{gilani_2021_hoopR}, we sampled one NBA starter from each team in the first game of each two season window, selecting players who appeared in that game and had at least 20 total games across the two season window. All player-game observations were restricted to games in which the player recorded non-zero minutes. Players were then followed across the two consecutive seasons, with forecasts estimated before each game. Players who were traded or missed games due to injury remained in the sample, as long as they met the minimum number of games criteria. We used seven windows, each covering two consecutive seasons, and avoiding the 2020 COVID season. The windows include 2010--11, 2012--13, 2014--15, 2016--17, 2018--19, 2023--24, and 2025--26. This yielded 207 players and approximately 22,000 observed Box Score outcomes.

\section{Results}

\begin{figure}[H]
    \centering
    \includegraphics[width=0.5\textwidth]{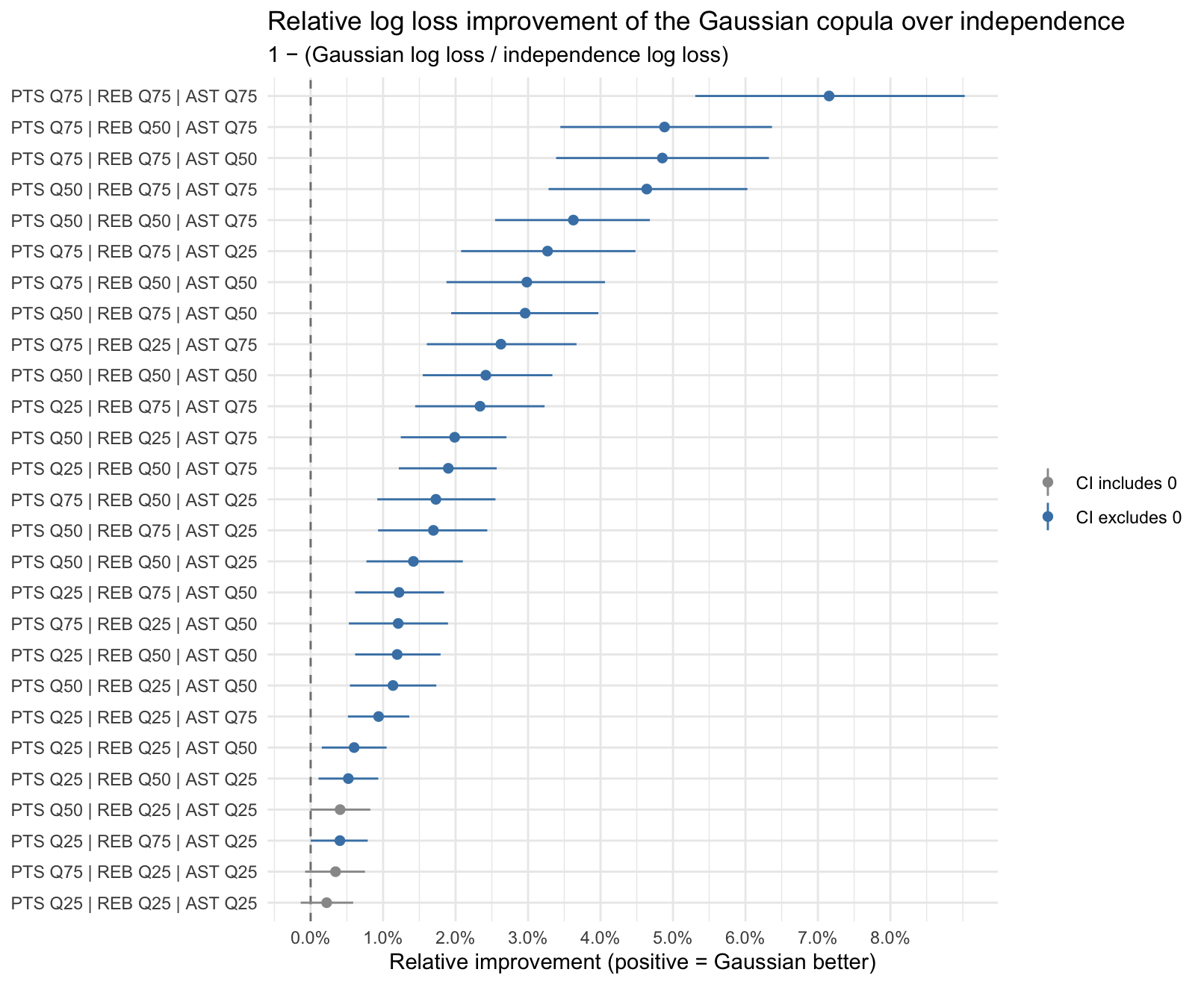}
    \caption{Relative Log Loss improvement of the Gaussian copula over the independence benchmark by threshold combination. Points show estimates and bars show confidence intervals.}
    \label{fig:relative-log-loss-independence}
\end{figure}

\begin{figure}[H]
    \centering
    \includegraphics[width=0.5\textwidth]{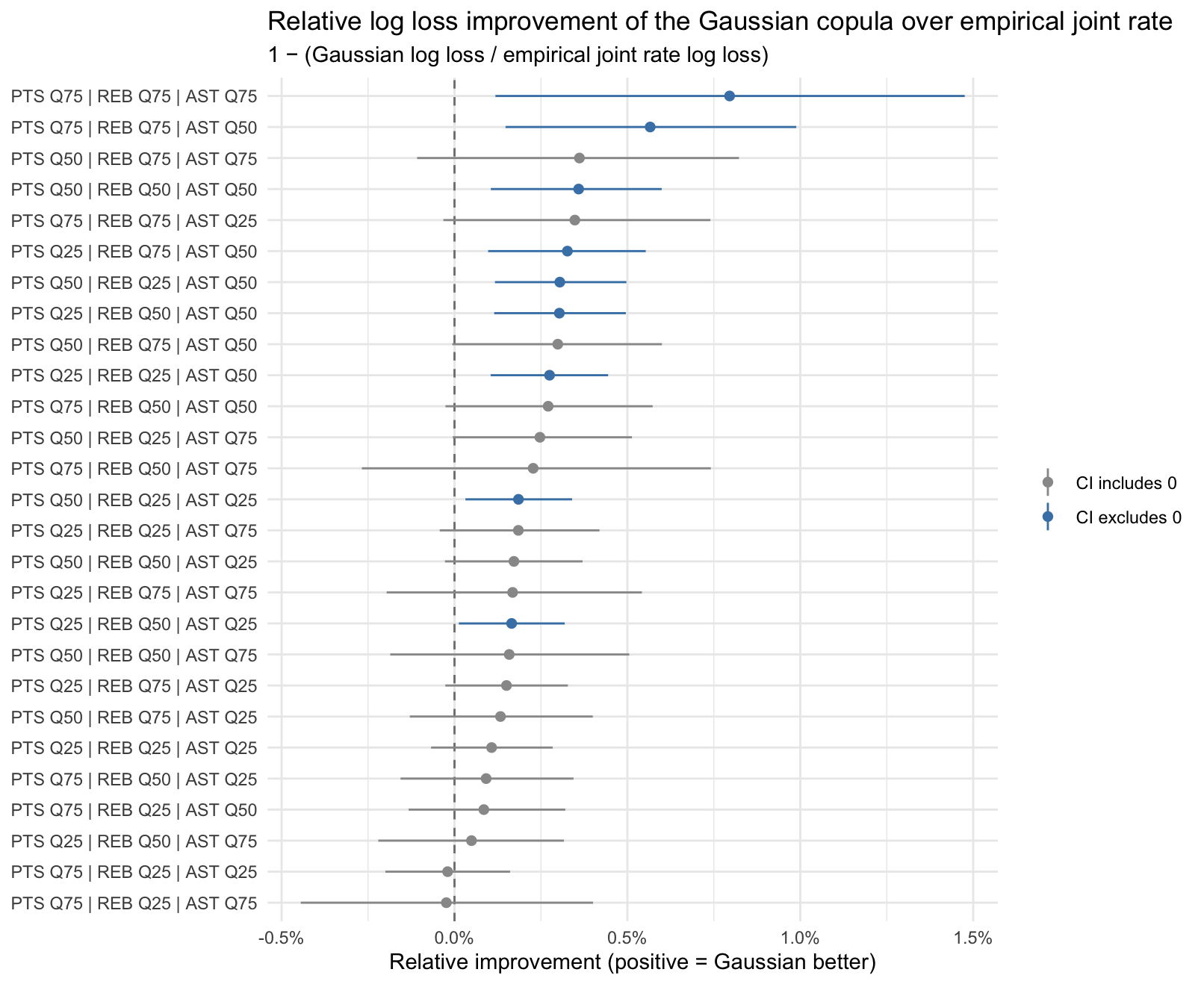}
    \caption{Relative log loss improvement of the Gaussian copula over the empirical benchmark by threshold combination. Points show estimates and bars show confidence intervals.}
    \label{fig:relative-log-loss-empirical}
\end{figure}

Figure~\ref{fig:relative-log-loss-independence} and~\ref{fig:relative-log-loss-empirical} compare the copula with each benchmark on the same
player-games. For each threshold combination, we report the relative improvement,
$1 - \ell_{\text{Gaussian}} / \ell_{\text{comparison}}$. Expressing the relative improvement makes it easier to compare across thresholds which have different underlying event probabilities. Because both models forecast the same
events, this is a paired comparison: player-games that are hard to forecast are hard for
both models, and this shared difficulty largely cancels in the ratio. Confidence intervals
come from a cluster bootstrap that resamples player-runs and recomputes both losses on each
resampled set, so they respect dependence among games from the same player. 

We observe clear separation between the Gaussian copula and the independence model with all but two thresholds showing statistically significant improvements. For thresholds with multiple Q75s we see between $2\%-8\%$ decreases in relative Log Loss, showing that the Gaussian copula is better at capturing the correlations at the upper quartiles.

The Log Loss looks very similar for the Gaussian copula and empirical model for many of the lower thresholds. Only at the upper thresholds do we see the Gaussian copula performing modestly better (roughly a $0.5\%$ improvement). We attach in the Appendix~\ref{sec:appendix} shows tables of Log Loss for more extreme thresholds which show greater outperformance. \\

For some context of the magnitude of Log Loss in Table~\ref{tab:log_loss_reference} we show the average Log Loss across example thresholds.

\begin{table}[H]
\centering
\caption{Average Log Loss Across Selected Threshold Combinations}
\label{tab:log_loss_reference}
\begin{tabular}{lrrr}
\toprule
Thresholds & Gaussian & Empirical & Independence \\
\midrule
Q25, Q25, Q25 & 0.651 & 0.651 & 0.652 \\
Q50, Q50, Q50 & 0.578 & 0.580 & 0.592 \\
Q75, Q75, Q75 & 0.249 & 0.251 & 0.269 \\
\bottomrule
\end{tabular}
\end{table}

At lower thresholds, we see very modest improvements over benchmarks, lower Log Loss is only clear at higher thresholds.

\subsection{Log Loss given Number of Training Games}

We bucket Log Loss improvements by the number of training games observed before that observation. This allows us to verify if any improvements observed by the Gaussian copula approach are limited to observations with small, or large, sample sizes.

\begin{figure}[H]
\centering
\begin{subfigure}[b]{0.48\textwidth}
    \centering
    \includegraphics[width=\textwidth]{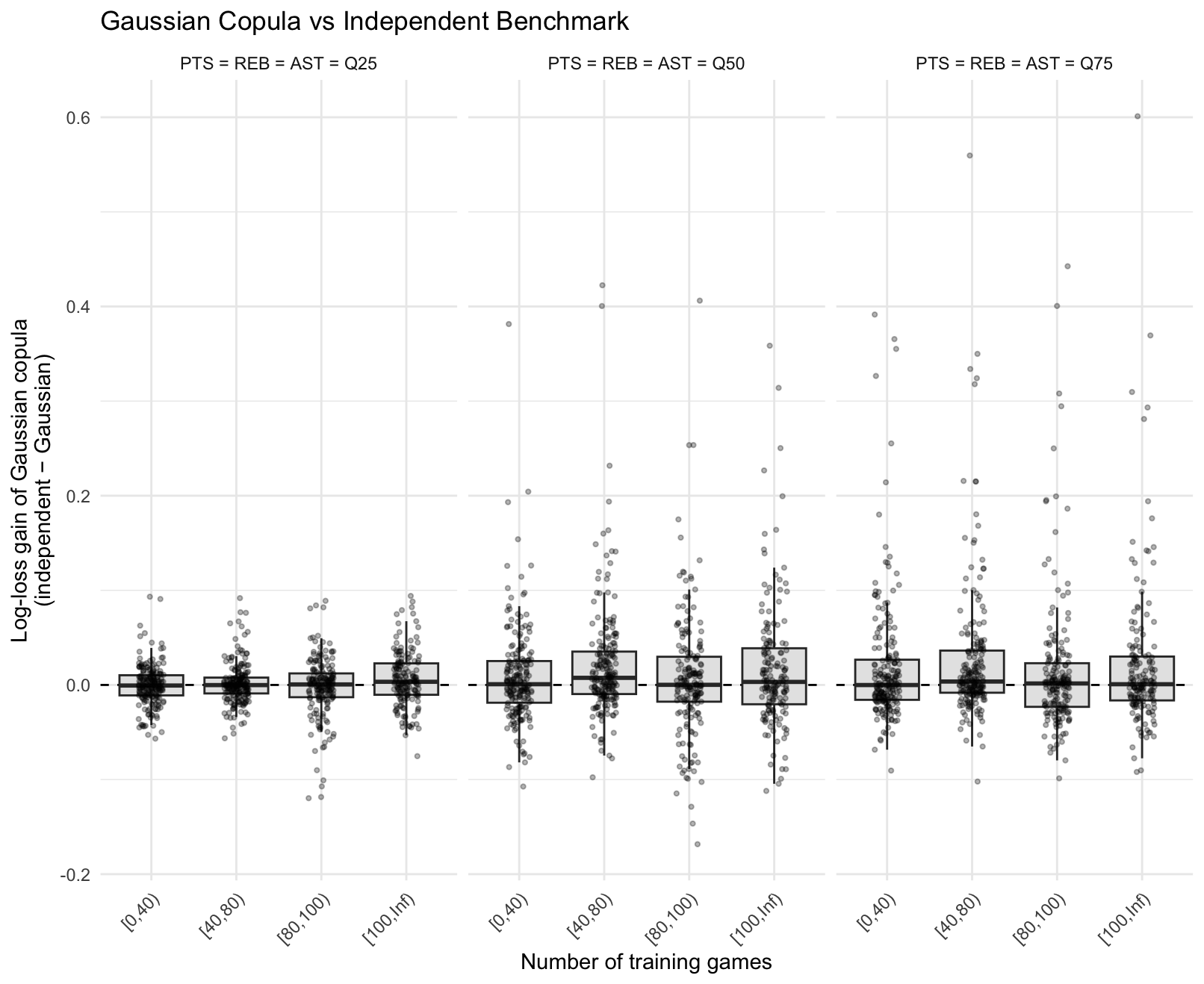}
    \caption{Independence}
    \label{fig:3d-left}
\end{subfigure}
\hfill
\begin{subfigure}[b]{0.48\textwidth}
    \centering
    \includegraphics[width=\textwidth]{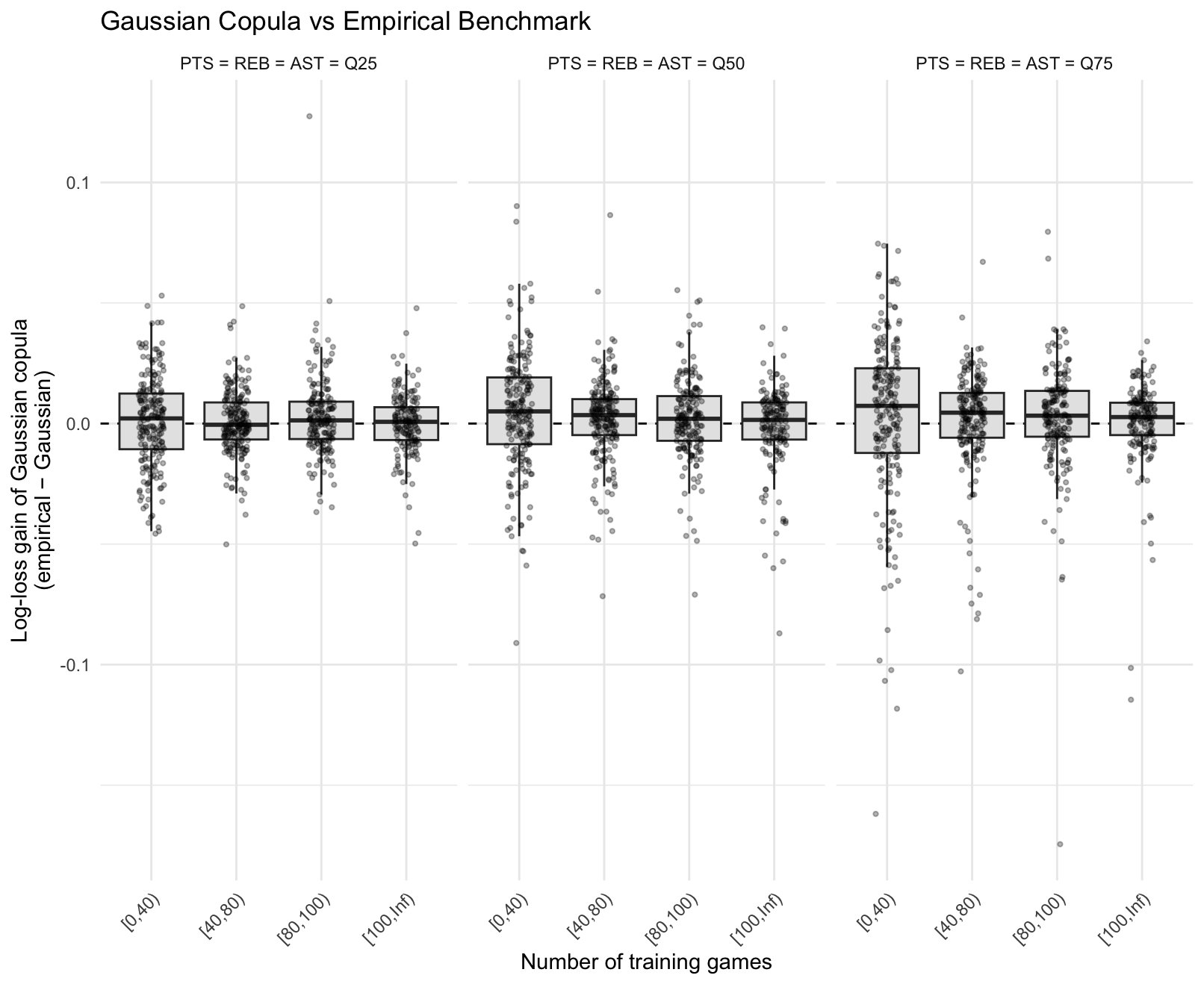}
    \caption{Empirical}
    \label{fig:3d-right}
\end{subfigure}
\caption{Log Loss bucketed by number of observed games}
\label{fig:bucket-games}
\end{figure}

We see little difference in outperformance of the Gaussian copula over the independence benchmark as we vary the number of observed training games, this seems to hold across thresholds. Conversely, for the empirical benchmark, as shown by the higher median, it appears that the Gaussian copula performs slightly better than the empirical method with fewer training games and this outperformance shrinks over time.

\subsection{Calibration Plots}

\begin{figure}[H]
  \centering
  \includegraphics[width=0.75\textwidth]{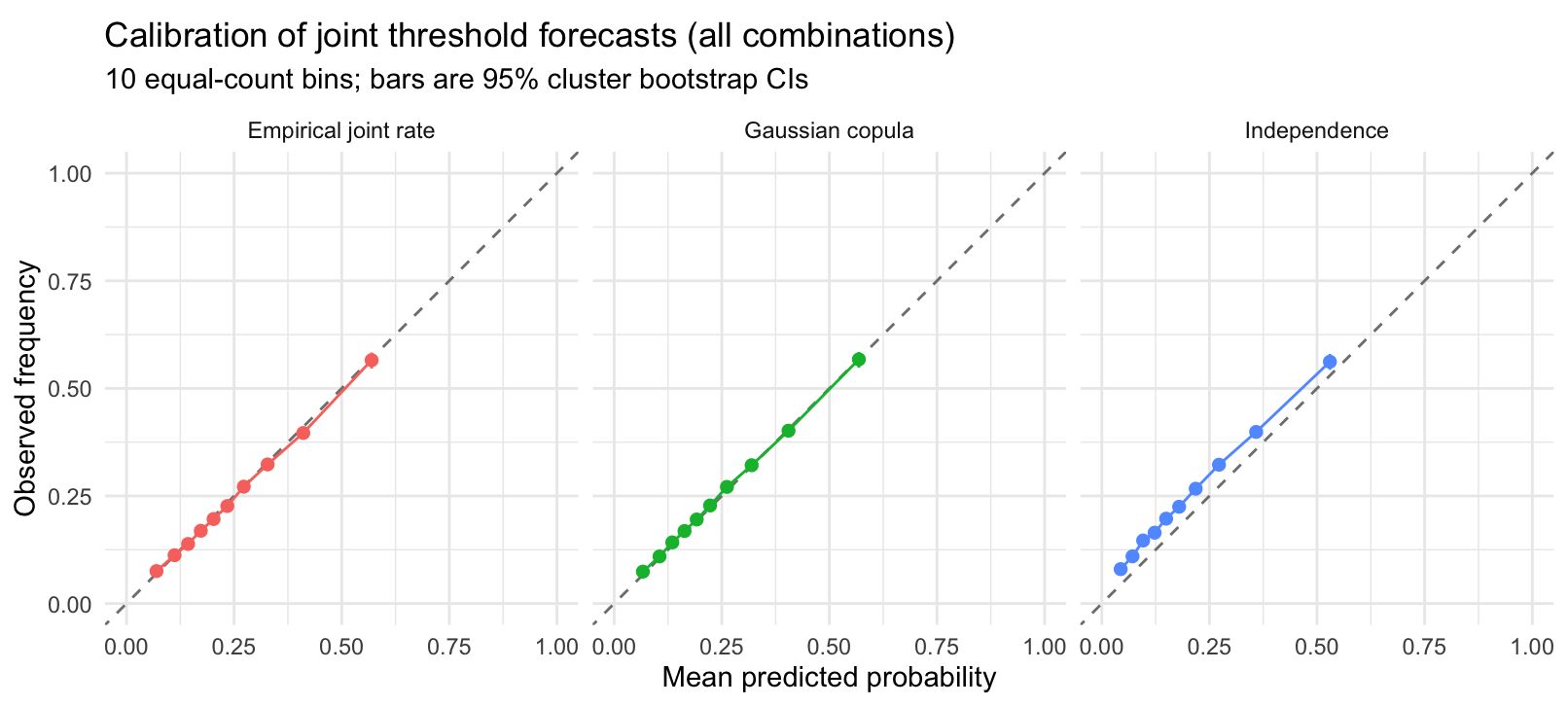}
  \caption{Calibration Plot for All Three Models}
  \label{fig:calib}
\end{figure}
We observe that the empirical and Gaussian models are well calibrated, being tightly clustered around the $45^\circ$ line, while the model assuming marginal independence systematically under predicts the observed frequency of joint exceedance.

\subsection{Relative Density Plots}

Figure~\ref{fig:3d-left} shows the ratio of the copula's joint exceedance probability to the independence benchmark, fixing assists at the Q75 threshold and varying the points and rebounds quartile levels. Figure~\ref{fig:3d-right} is the same graph but with assists fixed at Q25. We see in both cases the ratio of the probabilities rising as the quartile level increases, with a more extreme increase when assists are conditioned at Q75.

\begin{figure}[H]
\centering
\begin{subfigure}[b]{0.48\textwidth}
    \centering
    \includegraphics[width=\textwidth]{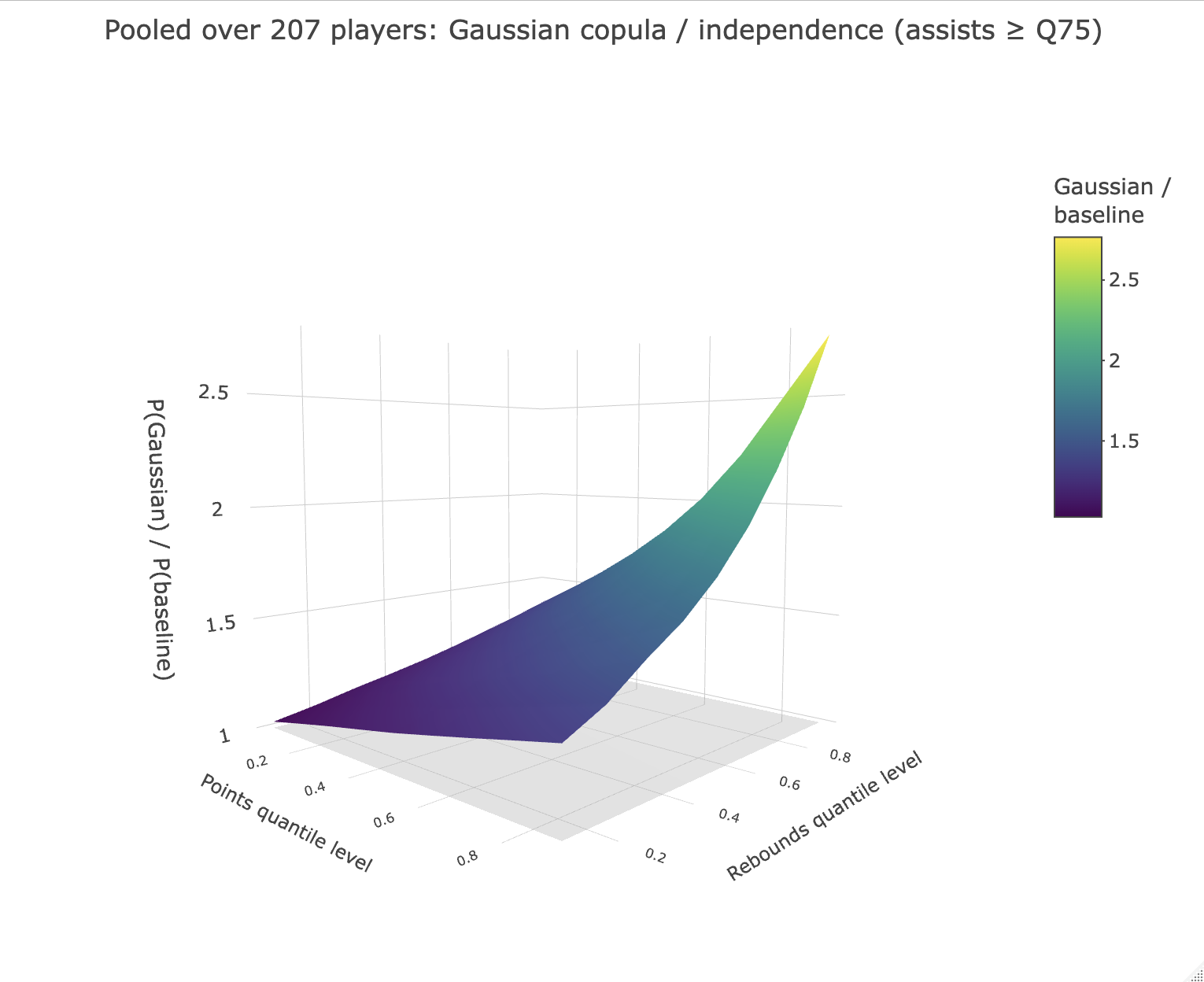}
    \caption{Independence benchmark comparison - Assists at Q75}
    \label{fig:3d-left}
\end{subfigure}
\hfill
\begin{subfigure}[b]{0.48\textwidth}
    \centering
    \includegraphics[width=\textwidth]{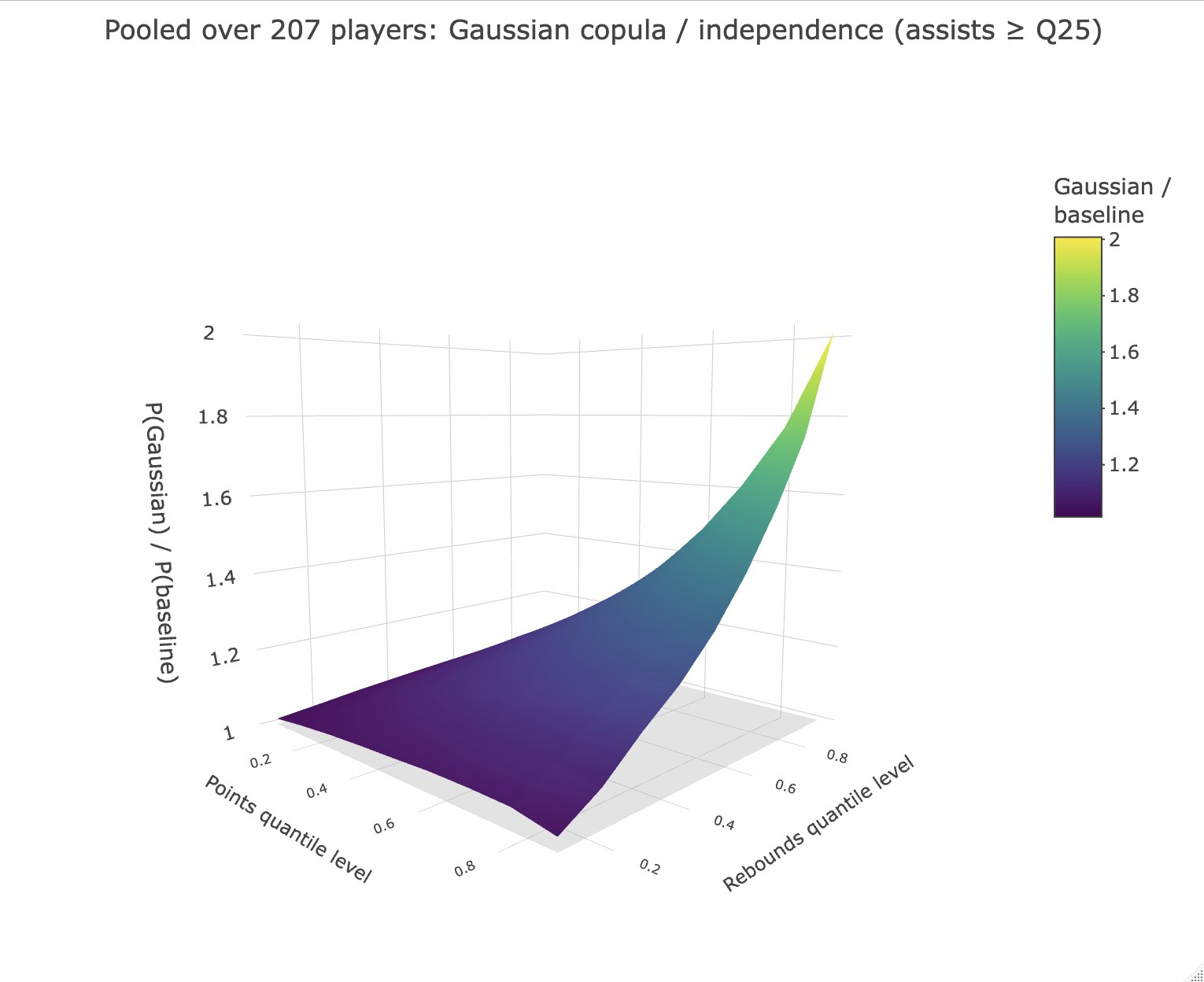}
    \caption{Independence benchmark comparison - Assists at Q25}
    \label{fig:3d-right}
\end{subfigure}
\caption{Ratio of the Gaussian copula's joint exceedance probability relative to the independence benchmark.}
\label{fig:3d-combined}
\end{figure}

Although the Gaussian distribution places relatively little probability mass in its tails, we observe that this does not prevent the Gaussian copula from capturing dependence at high quantiles. The copula separates the dependence structure from the marginal distributions, which in our case are empirical and therefore retain the observed tail behavior of each statistic.

\section{Conclusion}

In this paper we showed that treating the marginal likelihoods as independent is insufficient for predicting the joint likelihood of NBA Box Score statistics. A Gaussian copula performs better than the independence model immediately and more significantly at higher thresholds. It performs similarly to a Laplace smoothed empirical joint frequency model at lower threshold levels but performs modestly better at higher thresholds.
The result is a simple, fully generative model that can be used to calculate threshold probabilities of both observed and previously unobserved stat lines. It represents a baseline that future models could extend with richer pre-game context.

\section{Discussion and Future Directions}
\label{sec:discussion}

The results support modeling dependence: gains over independence are consistent across all 27 combinations, and gains over the empirical estimate are largest where joint events are rare and the empirical frequency is noisiest. Several limitations remain. The Gaussian copula imposes a symmetric dependence structure and exhibits zero asymptotic tail dependence. We explored alternative dependence specifications, including the Gumbel copula, as well as approaches that did not impose a parametric joint distribution and allowed the dependence structure to vary across players (Vine Copulas). We didn't find any clear improvement in predictive performance, so we retained the Gaussian copula as our primary model specification. 

Applications using Gaussian copulas such as \citep{li2000default} have been criticized for their lack of sensitivity to rare events and tail risk. However, this isn't particularly relevant in this situation since we're predicting binary outcomes, i.e. there aren't extreme asymmetric gains/losses to an incorrect prediction of an outlier.

We limited the scope of this paper to an individual's Box Score but the methodology could be expanded to assess the joint distribution of statistics across multiple players. However, this significantly expands the dimensionality of the problem, requiring the estimation of many dependencies, increasing the risk of identifying spurious relationships.

Our model uses only a player's own history, so it ignores ex ante knowledge of opponents, minutes, and pace. We suspect that model performance could be improved by conditioning on pre-game predicted possessions or by giving more weight to more recent observations. Finally, we evaluate player-relative quartile thresholds rather than the absolute lines offered by sports books, so we have not tested outperformance against market implied odds.

\bibliographystyle{apalike}
\bibliography{references}

\section{Appendix}
\label{sec:appendix}

All code and plots are accessible on \href{https://github.com/kennyws/NBA_Copula_Project}{GitHub}.

\subsection{LeBron James estimate}
\label{app:lebron}

$P(\text{PTS} \ge 27,\ \text{REB} \ge 7,\ \text{AST} \ge 7) \approx 6.5\%$ calculated on Nov 12th using this season and last season as sample.

\bigskip

We attach graphs of relative Log Loss improvement for more extreme thresholds.

\begin{figure}[H]
\centering
\begin{subfigure}[b]{0.48\textwidth}
    \centering
    \includegraphics[width=\textwidth]{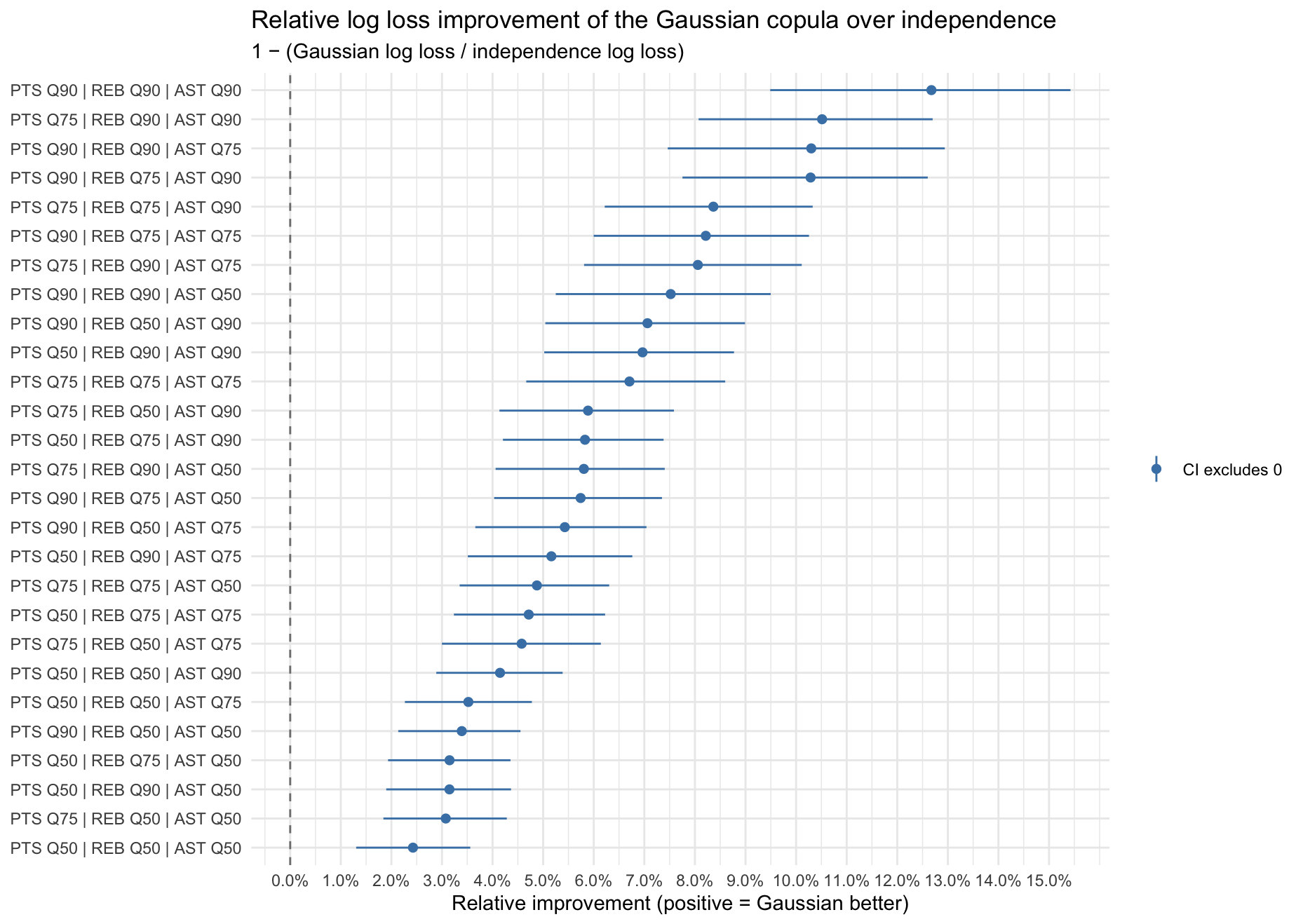}
    \caption{Gaussian vs Independence}
    \label{fig:High Threshold Independence}
\end{subfigure}
\hfill
\begin{subfigure}[b]{0.48\textwidth}
    \centering
    \includegraphics[width=\textwidth]{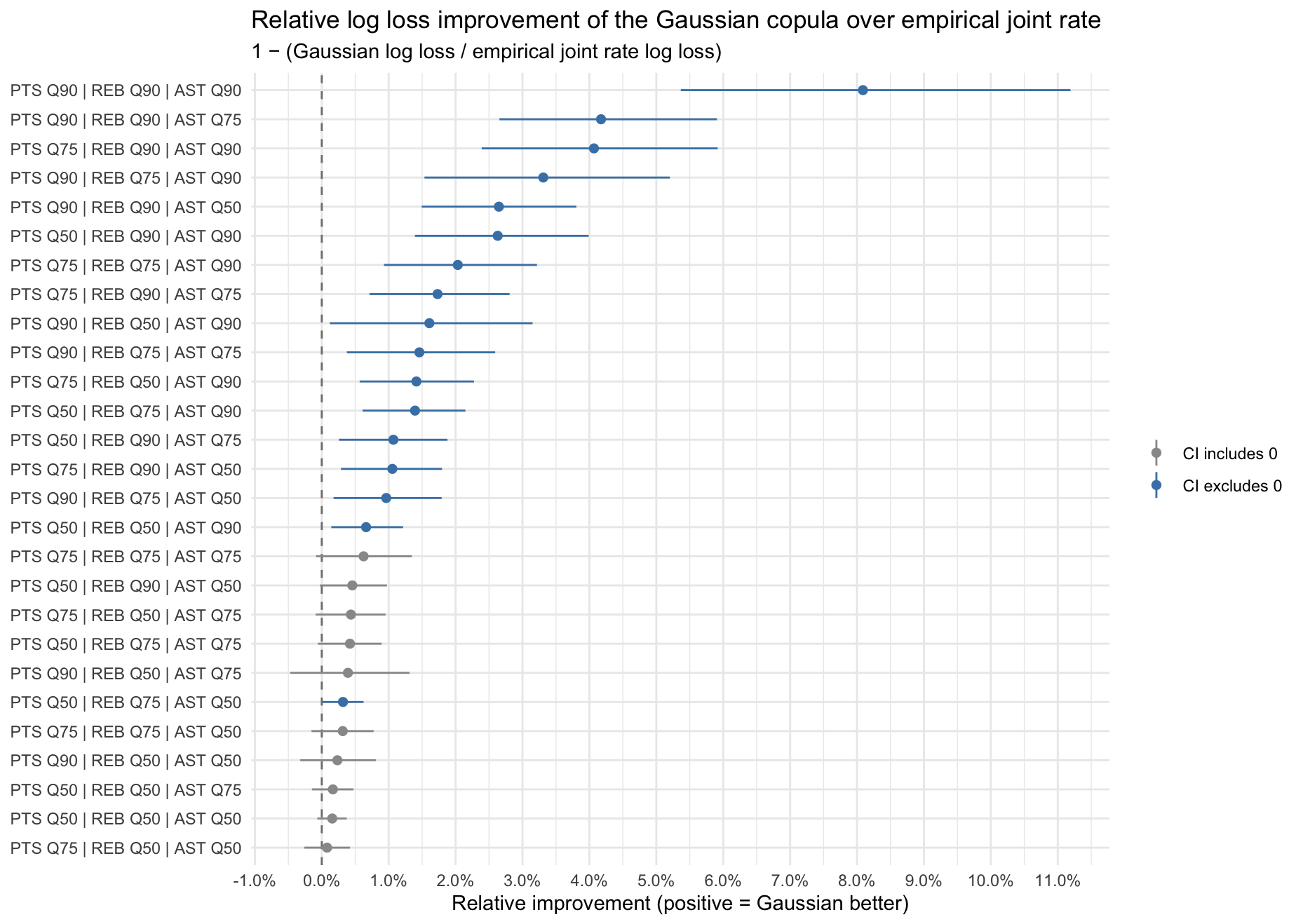}
    \caption{Gaussian vs Empirical}
    \label{fig:High Threshold Empirical}
\end{subfigure}
\caption{Relative Log Loss Improvement at High Thresholds}
\label{fig:Higher Thresholds}
\end{figure}

\newpage

Table~\ref{tab:independent_gaussian} and~\ref{tab:empirical_gaussian} show the mean paired differences in Log Loss between the Gaussian copula and alternative method, alongside bootstrapped p-values.

\begin{table}[H]
\centering
\caption{Independence vs. Gaussian Copula}
\label{tab:independent_gaussian}
\begin{tabular}{lrr}
\toprule
Combo & Mean Difference & $p_{\text{boot}}$ \\
\midrule
PTS Q75 | REB Q75 | AST Q75 & 0.01920 & 0.0001 \\
PTS Q50 | REB Q50 | AST Q75 & 0.01710 & 0.0001 \\
PTS Q75 | REB Q50 | AST Q75 & 0.01710 & 0.0001 \\
PTS Q75 | REB Q75 | AST Q50 & 0.01690 & 0.0001 \\
PTS Q50 | REB Q75 | AST Q75 & 0.01650 & 0.0001 \\
PTS Q50 | REB Q50 | AST Q50 & 0.01430 & 0.0001 \\
PTS Q50 | REB Q75 | AST Q50 & 0.01360 & 0.0001 \\
PTS Q75 | REB Q50 | AST Q50 & 0.01340 & 0.0001 \\
PTS Q75 | REB Q75 | AST Q25 & 0.01260 & 0.0001 \\
PTS Q50 | REB Q25 | AST Q75 & 0.01040 & 0.0001 \\
PTS Q25 | REB Q50 | AST Q75 & 0.01010 & 0.0001 \\
PTS Q75 | REB Q25 | AST Q75 & 0.01010 & 0.0001 \\
PTS Q25 | REB Q75 | AST Q75 & 0.00936 & 0.0001 \\
PTS Q50 | REB Q50 | AST Q25 & 0.00911 & 0.0001 \\
PTS Q75 | REB Q50 | AST Q25 & 0.00869 & 0.0004 \\
PTS Q50 | REB Q75 | AST Q25 & 0.00861 & 0.0001 \\
PTS Q25 | REB Q50 | AST Q50 & 0.00772 & 0.0001 \\
PTS Q50 | REB Q25 | AST Q50 & 0.00726 & 0.0002 \\
PTS Q25 | REB Q75 | AST Q50 & 0.00630 & 0.0002 \\
PTS Q75 | REB Q25 | AST Q50 & 0.00600 & 0.0008 \\
PTS Q25 | REB Q25 | AST Q75 & 0.00553 & 0.0001 \\
PTS Q25 | REB Q25 | AST Q50 & 0.00400 & 0.0090 \\
PTS Q25 | REB Q50 | AST Q25 & 0.00353 & 0.0136 \\
PTS Q50 | REB Q25 | AST Q25 & 0.00276 & 0.0560 \\
PTS Q25 | REB Q75 | AST Q25 & 0.00229 & 0.0454 \\
PTS Q75 | REB Q25 | AST Q25 & 0.00190 & 0.1020 \\
PTS Q25 | REB Q25 | AST Q25 & 0.00144 & 0.2330 \\
\bottomrule
\end{tabular}
\end{table}

\begin{table}[H]
\centering
\caption{Empirical vs. Gaussian Copula}
\label{tab:empirical_gaussian}
\begin{tabular}{lrr}
\toprule
Combo & Mean Difference & $p_{\text{boot}}$ \\
\midrule
PTS Q50 | REB Q50 | AST Q50 & 0.00208 & 0.0046 \\
PTS Q75 | REB Q75 | AST Q75 & 0.00200 & 0.0192 \\
PTS Q25 | REB Q50 | AST Q50 & 0.00194 & 0.0008 \\
PTS Q50 | REB Q25 | AST Q50 & 0.00193 & 0.0001 \\
PTS Q75 | REB Q75 | AST Q50 & 0.00188 & 0.0064 \\
PTS Q25 | REB Q25 | AST Q50 & 0.00183 & 0.0018 \\
PTS Q25 | REB Q75 | AST Q50 & 0.00167 & 0.0040 \\
PTS Q50 | REB Q75 | AST Q50 & 0.00134 & 0.0544 \\
PTS Q75 | REB Q75 | AST Q25 & 0.00130 & 0.0752 \\
PTS Q50 | REB Q25 | AST Q75 & 0.00127 & 0.0608 \\
PTS Q50 | REB Q25 | AST Q25 & 0.00125 & 0.0208 \\
PTS Q50 | REB Q75 | AST Q75 & 0.00123 & 0.1250 \\
PTS Q75 | REB Q50 | AST Q50 & 0.00119 & 0.0736 \\
PTS Q25 | REB Q50 | AST Q25 & 0.00112 & 0.0316 \\
PTS Q50 | REB Q50 | AST Q25 & 0.00109 & 0.1020 \\
PTS Q25 | REB Q25 | AST Q75 & 0.00108 & 0.1050 \\
PTS Q25 | REB Q75 | AST Q25 & 0.000853 & 0.0970 \\
PTS Q75 | REB Q50 | AST Q75 & 0.000758 & 0.3770 \\
PTS Q50 | REB Q50 | AST Q75 & 0.000721 & 0.3590 \\
PTS Q25 | REB Q25 | AST Q25 & 0.000699 & 0.2480 \\
PTS Q50 | REB Q75 | AST Q25 & 0.000665 & 0.3220 \\
PTS Q25 | REB Q75 | AST Q75 & 0.000659 & 0.3620 \\
PTS Q75 | REB Q50 | AST Q25 & 0.000452 & 0.4650 \\
PTS Q75 | REB Q25 | AST Q50 & 0.000417 & 0.4730 \\
PTS Q25 | REB Q50 | AST Q75 & 0.000258 & 0.7060 \\
PTS Q75 | REB Q25 | AST Q75 & -0.000088 & 0.9310 \\
PTS Q75 | REB Q25 | AST Q25 & -0.000110 & 0.8220 \\
\bottomrule
\end{tabular}
\end{table}

\end{document}